\documentclass[preprint,aps,12pt,preprintnumbers,eqsecnum,nofootinbib]{revtex4}
\usepackage{graphicx}
\usepackage{subfigure}

\usepackage{color}
\usepackage{amssymb,amsmath}

\newcommand{\beq}{\begin{equation}}
\newcommand{\enq}{\end{equation}}

\begin{document}
%
%
% Title of paper
\title{\vspace*{0.5in} 
Model Sensitivity in Holographic Superconductors and their Deconstructed Cousins
\vskip 0.1in}
\author{Joshua Erlich and Zhen Wang}

\affiliation{High Energy Theory Group, Department of Physics,
College of William and Mary, Williamsburg, VA 23187-8795}
\date{\today}
\begin{abstract}
Holographic models of superconductors successfully reproduce certain experimental features of
high-temperature superconductors, such as a large gap-to-$T_{\rm c}$ ratio compared to that of conventional superconductors.  By deconstructing the extra dimension of  these holographic models, similar phenomenology is described by a class of models defined in the natural dimension of the superconducting system.  We analyze the sensitivity of certain observables in holographic and deconstructed holographic superconductors to details of the extra-dimensional spacetime.  Our results support the notion that certain quantitative successes of simple models of this type are accidental.  However, we also find a certain universal relationship between superconducting observables.

\end{abstract}
\maketitle

\section{Introduction}
Holographic models of nonperturbative physical systems have been more successful quantitatively than should have been expected.  The  most developed applications of holographic model building are to quantum chromodynamics (QCD) \cite{holoqcd,Sakai-Sugimoto}, electroweak symmetry breaking \cite{holo-ewsb} and condensed matter systems, especially superconductors \cite{Gubser,HHH}.   Arguments based on insensitivity to model
details \cite{adsqcd-universality}, approximate conformal invariance \cite{conformal} and decoupling of high-dimension states and operators \cite{highdim} have been put forward in an attempt to understand the unreasonable effectiveness of some of these models.  

Holographic models of 3+1 dimensional systems are 4+1 dimensional theories in which the behavior of fields near the boundary of the spacetime, typically Anti-de Sitter (AdS) space, determines the properties of the corresponding lower-dimensional system.  However, gauge theories in more than 3+1 dimensions are generally nonrenormalizable.  Deconstruction of extra dimensions provides a gauge-invariant completion  of  higher-dimensional gauge theories \cite{HPW,deconstruction}.  A deconstructed extra-dimensional model is a lower-dimensional theory which, below some energy scale, has an effective description in which one or more extra dimensions are latticized.  Deconstruction is useful for model building in that it is sometimes possible to reduce the number of ``lattice sites'' to just a few while maintaining the interesting phenomenology of a higher-dimensional model, yielding a relatively simple model of the system of interest.  For example, in the context of electroweak symmetry breaking, deconstruction provides one route to little Higgs models \cite{little-Higgs}.  More recently, this approach has been used to construct models with some of the properties of holographic superconductors, even though defined in the natural dimension of the superconducting system \cite{ACE}.

Among the successful predictions of holographic models  are certain features of high-temperature superconductors such as an enhanced ratio of the superconducting gap $(\Delta)$ to the critical temperature $(T_{\rm c})$ \cite{HHH}.  Bottom-up holographic models of finite-temperature systems  typically begin with an AdS-Schwarzschild or AdS-Reissner-Nordstrom black hole geometry, the latter taking into account the backreaction of charge density on the geometry.  These geometries are chosen mainly for simplicity, but in holographic models of superconductors derived from string theory \cite{Gubser}, the spacetime geometries may be more complicated and depend on the fluxes of fields associated with  D-brane configurations.  Other geometric backgrounds in holographic models arise as the induced metric on a brane embedded in a higher-dimensional spacetime, such as on the flavor branes in the  holographic QCD model of Sakai and Sugimoto \cite{Sakai-Sugimoto}, and these induced geometries are not derived as the solution to Einstein's equation with a specified energy-momentum tensor.  

It is the goal of this paper to explore the sensitivity of observables to variations in the details of holographic models of superconductors and  in deconstructed variations of those models. As such, we consider holographic superconductors in generalizations of the 3+1 dimensional AdS-Schwarzschild metric.  
We find certain generic features in the phenomenology of these models, but details such as the ratio of the superconducting gap to the critical temperature are sensitive to the model details, which suggests that successful quantitative predictions in prototypical models are likely accidental.  This is not to say that those models will not prove valuable in explaining the puzzling properties of unconventional superconductors, only that quantitative predictions are more model dependent than one might have hoped.  It has already been noted that there are quantitative and even qualitative distinctions between superconducting models, for example between those which take into account the backreaction of the charge density on the metric and those that don't \cite{HHH}.  The work presented here focuses on sensitivity to the extra-dimensional spacetime, parametrizing the AdS black-hole metric in a particular way in order to quantify the variability of superconducting observables in a class of holographic models and in deconstructed versions of those models.

\subsection{Holographic Superconductors}
Here we briefly review the construction of holographic superconductor models and the calculation of observables in those models.  In a holographic superconductor, a charged field condenses in an extra-dimensional black-hole background whose Hawking temperature is below some critical temperature $T_{\rm c}$.  The temperature of the lower-dimensional system is identified with the Hawking temperature of the higher-dimensional black hole \cite{Witten-holoconfinement}. The charged
condensate spontaneously breaks the electromagnetic U(1) gauge group\footnote{To be precise,
the U(1) gauge invariance of the holographic model corresponds to a global U(1)
symmetry of the lower-dimensional  system.  However, as argued in Ref.~\cite{HHH}, this global U(1) can be weakly gauged in
order to determine some aspects of the dynamics of the corresponding superconducting system.}  and gives rise to  superconducting phenomenology \cite{HHH}.

In this work we ignore the backreaction of the charge density on the spacetime geometry, and for now we consider an Abelian Higgs model in a 3+1 dimensional AdS-Schwarzschild spacetime background.  This is meant to describe a system which is superconducting in two spatial dimensions, {\em e.g.} the copper-oxide planes of cuprate superconductors. 
We can choose coordinates such that the lengths are normalized to the AdS scale and the metric has the form \begin{equation}
ds^2=\frac{1}{z^2}\left[f(z)dt^2-\frac{1}{f(z)}dz^2-(dx^2+dy^2)\right],
\label{eq:metric}
\end{equation}
where 
\begin{equation}
f(z)=1-\frac{z^p}{z_H^p}, \label{eq:f}\end{equation}
with $p=3$ corresponding to the 3+1 dimensional AdS-Schwarzschild metric.

With the Euclidean time $\tau\equiv it$ compactified with period $1/T$, the Hawking temperature associated with the modified black-hole metric follows from the condition that there be no conical singularity at the horizon.  In the absence of a conical singularity, if $z^*$ is the proper distance from the horizon $z=z_H$ to a nearby point displaced only in the radial ($z$) direction and $\beta^*$ is the proper circumference of the Euclidean-time circle at that fixed radial position, then $2 \pi z^{*} = \beta^*$. For metrics of the form (\ref{eq:metric}), for small proper displacements from the horizon,
\begin{eqnarray}
2\pi z^*&=&2 \pi \int_{z_H-\varepsilon}^{z_H} \frac{dz}{z \sqrt{f(z)}} = 
\int_{z_H-\varepsilon}^{z_H}\frac{dz}{z_H\sqrt{-(z_H-z)f'(z_H)}} \nonumber \\
&=&\frac{4\pi\sqrt{\varepsilon}}{z_H\sqrt{-f'(z_H)}}, \\
\beta^*&=&\frac{\sqrt{f(z_H-\varepsilon)}}{z_H-\varepsilon}  \frac{1}{T} 
=\frac{\sqrt{-f'(z_H)}}{z_H}  \frac{\sqrt{\varepsilon}}{T} \,,
\label{eq:temp}
\end{eqnarray}
and with $f(z)$ given by Eq.~(\ref{eq:f}), the Hawking temperature is then:
\begin{equation}
T =-\frac{4\pi}{f'(z_H)} =\frac{p}{4 \pi z_{H}} \, .
\end{equation}

In the continuum model, observables are independent of the choice of coordinates.  However, away from the continuum limit, the deconstructed models are sensitive to the latticization of the extra
dimension, which in turn depends on the coordinate choice.  
In the continuum model, the action for the scalar field $\psi$ and U(1) gauge field $A_M$ ($M\in \{0,1,2,3\}$) is \begin{equation}
S=\int d^4x\,\sqrt{g}\left[ -\frac{1}{4}F_{MN}F^{MN}+\left|\left(\partial_M-iA_M\right)\psi\right|^2
-m^2 |\psi|^2\right],
\end{equation}
where $g_{MN}$ is the metric defined by Eq.~(\ref{eq:metric}).  For definiteness we take $m^2=-2$ in AdS units, as in Refs.~\cite{HHH} and \cite{ACE}.

Near the boundary $z=0$, the field $\psi$ has solutions \begin{equation}
\psi(z)\sim\psi^{(1)}z + \psi^{(2)}z^2.\label{eq:psisol}\end{equation}
In this model both independent solutions for $\psi(x,z)$ have finite action, so the AdS/CFT interpretation of the
two solutions is ambiguous.  Here we choose the interpretation that $\psi^{(2)}$ is the condensate
of the Cooper pair operator, while $\psi^{(1)}$ would then be the external source for that operator, which we assume vanishes.  Hence, $\psi^{(1)}=0$ is a boundary condition for the solutions of interest.

The bulk U(1) gauge field, $A_M$, is dual to the electric current and the background electromagnetic field.  In order to allow for nonvanishing chemical potential and charge density, we consider solutions in which the time component, $A_0$, is nonvanishing.  The equations of motion
have solutions for which $A_0$  behaves near the boundary as \begin{equation}
A_0\sim \mu - \rho z, \label{eq:Asol}\end{equation}
where $\mu$ is the coefficient of the non-normalizable solution and is identified with the chemical potential, which is a source for $\rho$, the charge density.

The phenomenology of the model is determined by fixing the temperature $T$ as it appears in the AdS black-hole metric, solving the coupled equations of motion
for $\psi$ and $A_M$ subject to the ultraviolet ({\em i.e.} $z=0$) boundary conditions $\psi^{(1)}=0$, $A_0(0)=\mu$, and the infrared ({\em i.e.} $z\rightarrow z_H$) boundary conditions $A_0(z_H)=0$ and $f'(z_H)z_H\psi'(z_H)=m^2\psi(z_H)$.  The last condition follows from the equations of motion, but is enforced as a regularity condition on the numerical solutions.  
The Cooper pair condensate $\langle O_2\rangle$ and background charge density are then determined by $\psi^{(2)}$ ({\em c.f.} Eq.~(\ref{eq:psisol})) and $\rho$ ({\em c.f.} Eq.~(\ref{eq:Asol})), respectively.  Varying the temperature $T$ then allows for a determination of the phase structure of the model, as $\langle O_2\rangle= 0$ for $T> T_{\rm c}$.

To analyze the frequency-dependent conductivity we fix the background for $\psi$ and instead solve the equations of motion for $A_M$ in a background with $A_a= e^{-i\omega t}\varepsilon_aA(z)$, $a\in\{1,2\}$, corresponding to a uniform oscillating background electric field $E_a=\partial_0 A_a|_{z\rightarrow0}$, polarized in the $\varepsilon_a$ direction.  Solutions are chosen to be ingoing at the horizon in order to enforce causal behavior of the current two-point function 
\cite{son-starinets}. The solution for $A_a\sim A_a^{(0)}+J_a z$ as $z\rightarrow0$ then determines the electric current $J_a(\omega)$, from which the conductivity, $\sigma=J^a/E_a$ follows.  A generic
feature of superconductors is the existence of a frequency gap $\omega_g$
below which there
are no modes available to excite and generate a current, 
so that $\sigma(\omega)=0$ for $\omega<\omega_g$  for $T=0$.  For nonvanishing temperature, even as $\omega\rightarrow0$ the current may be nonvanishing, where for small enough temperature  $\sigma(\omega\rightarrow0)\propto\exp^{-\Delta/T}$, where $\Delta$ is the superconducting gap.  From the weakly coupled BCS theory, we would expect $\Delta\approx\omega_g/2$, which also appears to be satisfied in the original model of Ref.~\cite{HHH}.

\subsection{Deconstructed Holographic Superconductors}
We will study a class of models, based on the models of Ref.~\cite{ACE}, in which the extra dimension of the holographic superconductor is deconstructed.  Models with certain similarities to these were also considered in Ref.~\cite{Kiritsis}.  The higher-dimensional U(1) gauge theory is replaced by a U(1)$^N$ gauge theory in one fewer dimension, where $N\rightarrow\infty$ in the continuum limit.  Scalar link fields charged under ``neighboring'' pairs of U(1) gauge groups are arranged to have prescribed expectation values, breaking the U(1)$^N$ gauge group in such a way that the resulting action is that of the latticized higher-dimensional theory.  The massive gauge fields replace the Kaluza-Klein modes in the continuum model.  The fluctuations of the link fields do not correspond to degrees of freedom in the continuum theory, so we assume that they are heavy compared to the scales of interest in our analysis and disregard them in our analysis.

Expanding the fields in components, the action of the holographic model is, \begin{eqnarray}
S&=&\int d^4x\left[\frac{1}{2}F_{0z}^2+\frac{1}{2f(z)}F_{0a}^2-\frac{f(z)}{2}F_{za}^2-\frac{1}{4}F_{ab}^2+\frac{1}{z^2f(z)}\left|\partial_0\psi-iA_0\psi\right|^2 \right.\nonumber \\
&&\left.   -\frac{f(z)}{z^2} \left|\partial_z \psi -i A_z \psi \right|^2-\frac{1}{z^2} \left|\partial_i \psi -i A_i \psi\right|^2 
- \frac{1}{z^4} m^2 |\psi|^2 \right] \, ,
\end{eqnarray}
where the lower-case Latin indices $a,b$ are summed over the $x$ and $y$ coordinates.  We now
latticize the spacetime in one dimension by replacing the $z$ coordinate by a discrete set of $N$ points:
\begin{equation}
z_j = \left\{ \begin{array}{ll}
\epsilon+(j-1)a  & {\rm for}\  j=1, \ldots, N-1 \\
\epsilon+(N-2)\, a + a_H & {\rm for}\ j=N \,\,\, , \end{array}\right. \,
\end{equation}
where $z_N = z_H$, $a$ is the lattice spacing in $z$-coordinates, and $\epsilon$  is a UV  cutoff.  The Lagrangian for the deconstructed theory is of the form,
\begin{equation}
{\cal L}=  \sum_{j=2}^{N-1} \left[- \frac{1}{4}(F_{\mu\nu})_j (F^{\mu\nu})_j
+Z_j | D_\mu \psi_j |^2 \right]+\sum_{j=1}^{N-1} \left[ |D_\mu \Sigma_j |^2 - Z_j V_j \right],
\end{equation}
where $V_j$ is the scalar potential for link field $\Sigma_j$, and the coefficients $Z_j$ and
metric factors $g_{\mu\nu}^j$ by which indices are contracted vary with lattice position $j$. 
The parameters in the model may be chosen (see Ref.~\cite{ACE} for more details) such that the effective theory below the scale set by the link fields is given by the Lagrangian, \begin{eqnarray}
 {\cal L} &=& \sum _{j=1}^{N-1} a_j   \left[ \frac{1}{2}(\phi'_j)^2 -\frac{f_j}{2} ({A'_a}_j)^2 -\frac{f_j}{z_j^2} |\psi'_j |^2 \right]
+\sum _{j=2}^{N-1} a_j \left[ \frac{1}{2 f_j} (F_{0a})_j^2  -\frac{1}{4} (F_{ab})_j^2 \right] \nonumber \\
&+& \sum _{j=2}^{N-1} a_j  \left[ \frac{1}{z_j^2 f_j} | \partial_0 \psi_j -i {\phi}_j \psi_j |^2
-\frac{1}{z_j^2} |\partial_a \psi_j -i {A_a}_j \psi_j|^2 - \frac{1}{z_j^4} m^2 |\psi_j|^2 \right] \, ,
\end{eqnarray}
where $\phi_j\equiv A^0_j$,  $
f_j\equiv f(z_j)$, and the primes correspond to discretized derivatives, for example
\begin{equation}
\phi_j'\equiv \frac{\phi_{j+1}-\phi_j}{a}. \label{eq:deriv}\end{equation}

The U(1) gauge group at the first lattice site (the UV boundary site) is identified with the electromagnetic gauge group.  Solutions to the equations of motion with discretized versions of the boundary conditions on the fields $\psi$ and $A_M$ allow for the calculation of observables by analogy with the holographic analysis in the continuum model \cite{ACE}.  In the case of a small number of lattice sites there is no {\em a priori} reason to expect phenomenology similar to that of the continuum model.  Indeed, we find significant deviation from the predictions of the continuum model, though certain qualitative features remain.
More complete details of these computations are presented below.  

\section{Results} \label{sec:results}

To consider the sensitivity of observables to the  spacetime geometry, we allow
the power $p$ in Eq.~(\ref{eq:f}) to deviate from its value $p=3$ corresponding to the AdS-Schwarzschild spacetime.
For generic $p$ the metric is not a solution to Einstein's equations with a prescribed energy-momentum tensor.  However, the initial choice of AdS-Schwarzschild geometry was made for simplicity and is equally arbitrary, and we can imagine either fluxes of fields that would give rise to the requisite energy-momentum tensor, or we can imagine that the class of spacetimes described here corresponds to the induced metric on a brane embedded in a particular higher-dimensional spacetime.  The goal here is to parametrize a class of deviations from the prototypical spacetime in order to analyze the sensitivity of observables to the detailed form of the spacetime metric.  Our particular choice of parametrized metric is mostly arbitrary, though  the class of spacetimes considered here remains asymptotically AdS near the boundary at $z=0$.

\begin{figure}[ht]
\subfigure{\includegraphics[width = 0.495\textwidth]{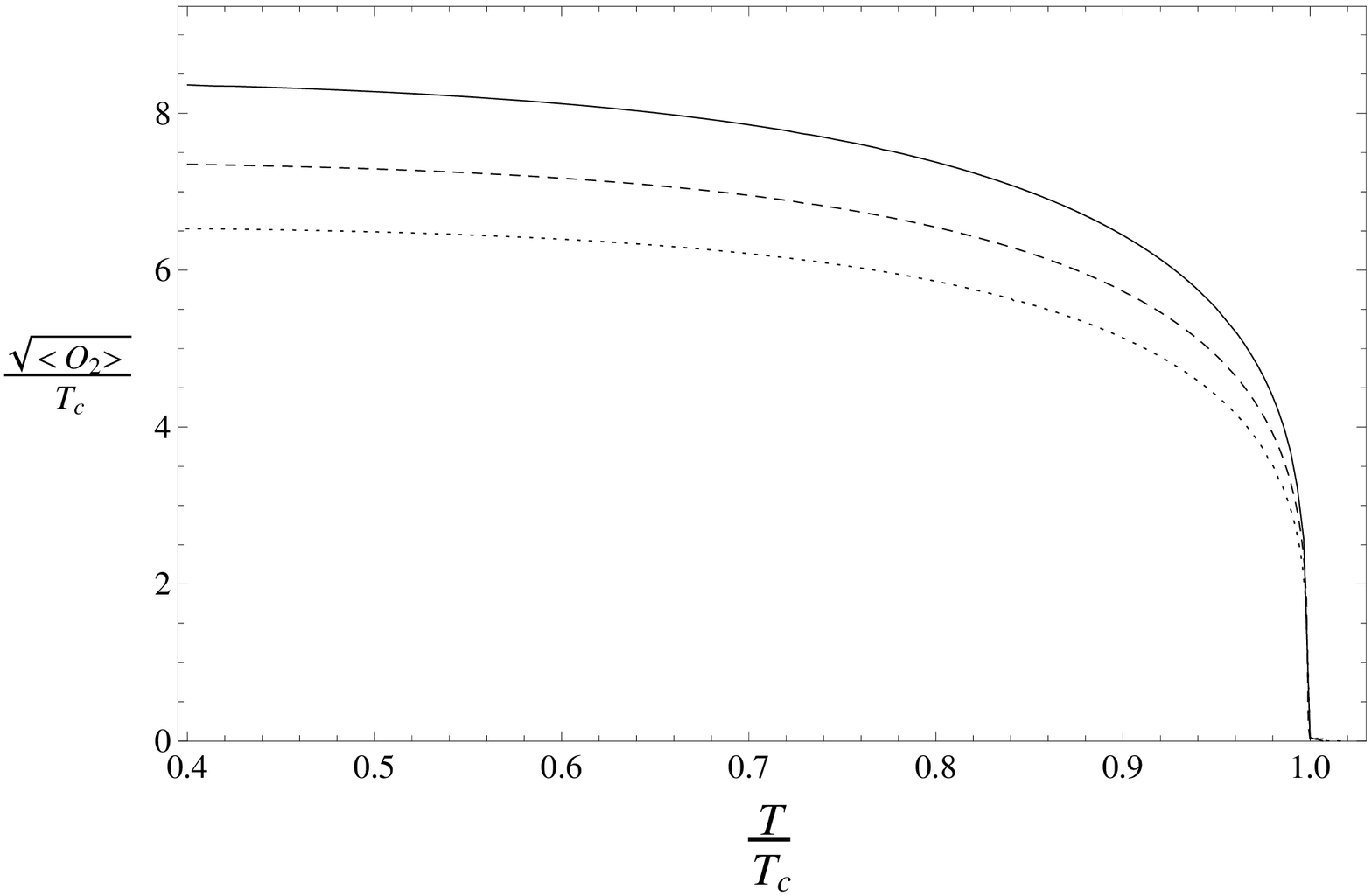}}
\subfigure{\includegraphics[width = 0.495\textwidth]{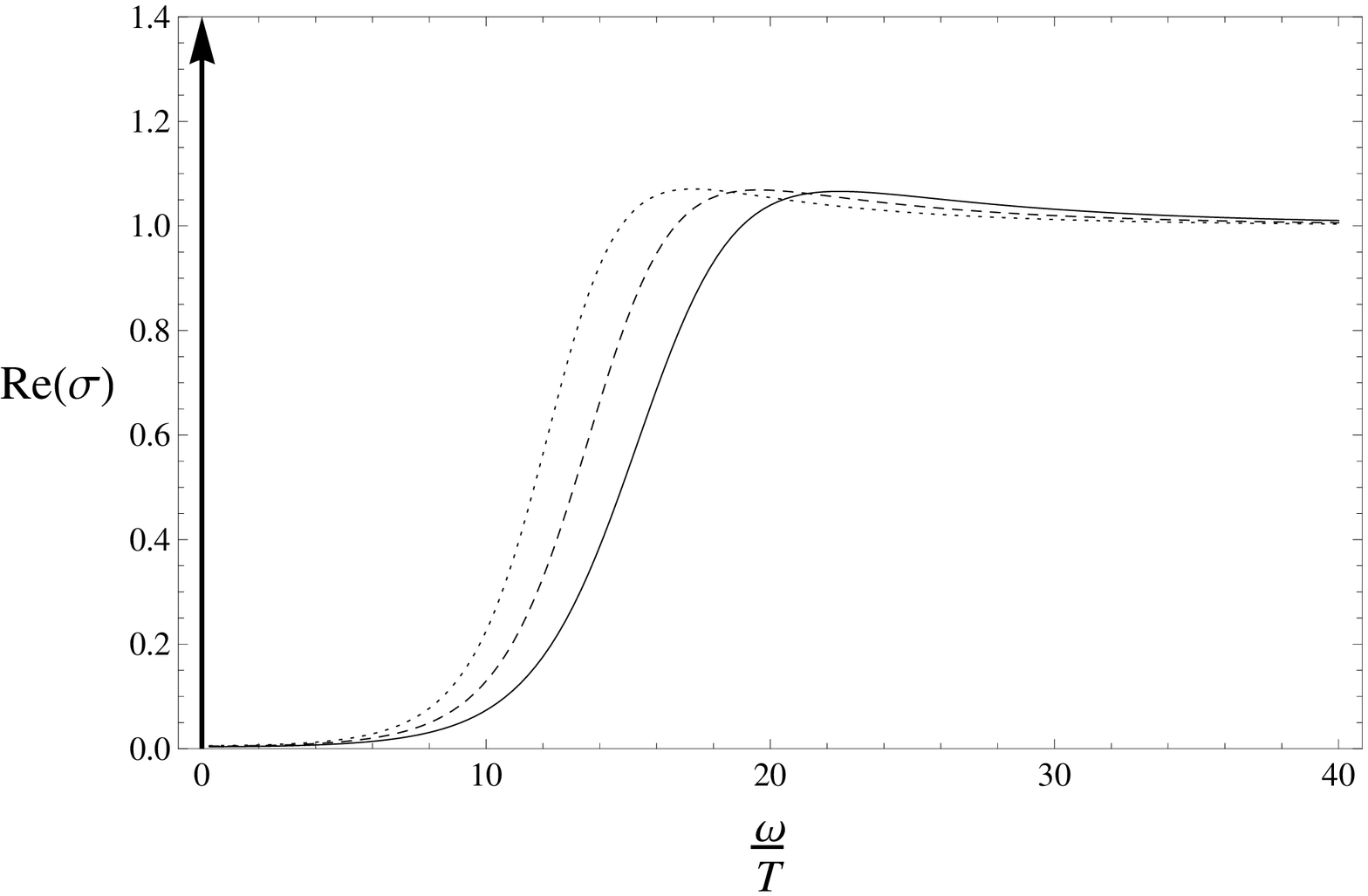}}
 \caption{
The condensate  $\langle O_{2} \rangle$ and the real part of conductivity $\sigma$, at $T/T_{\rm c} = 0.5$, for the continuum theory. Solid, dashed and dotted curves have $p = 3, 3.5$ and $4$, respectively. The arrow indicates a Dirac delta function. The critical temperature $T_{\rm c}$ in unit of $\rho^{1/2}$ for $p = 3, 3.5$ and $4$ are $0.119, 0.135$ and $0.153$, respectively.}
\label{fig:contin}
\end{figure}

We first analyze the continuum theory for $p = 3, 3.5$ and $4$. In our numerics, we cut off the spacetime in the  UV   at $z=10^{-4}$ and near the horizon at  $z=z_H-10^{-5}$. 
We impose  the boundary conditions discussed previously, and we fix $\rho=1$, which  by a scaling relation in the model also fixes $T_{\rm c}\propto\rho^{1/2}$ 
\cite{HHH}.
The superconducting condensate  and the real part of the conductivity are plotted in Fig.~\ref{fig:contin}.  The delta function in the real part of the conductivity may be inferred from a pole 
in the imaginary part, not shown in the figure, by the Kramers-Kronig relation for the conductivity.

\begin{figure}
\subfigure{\includegraphics[width = 0.32\textwidth]{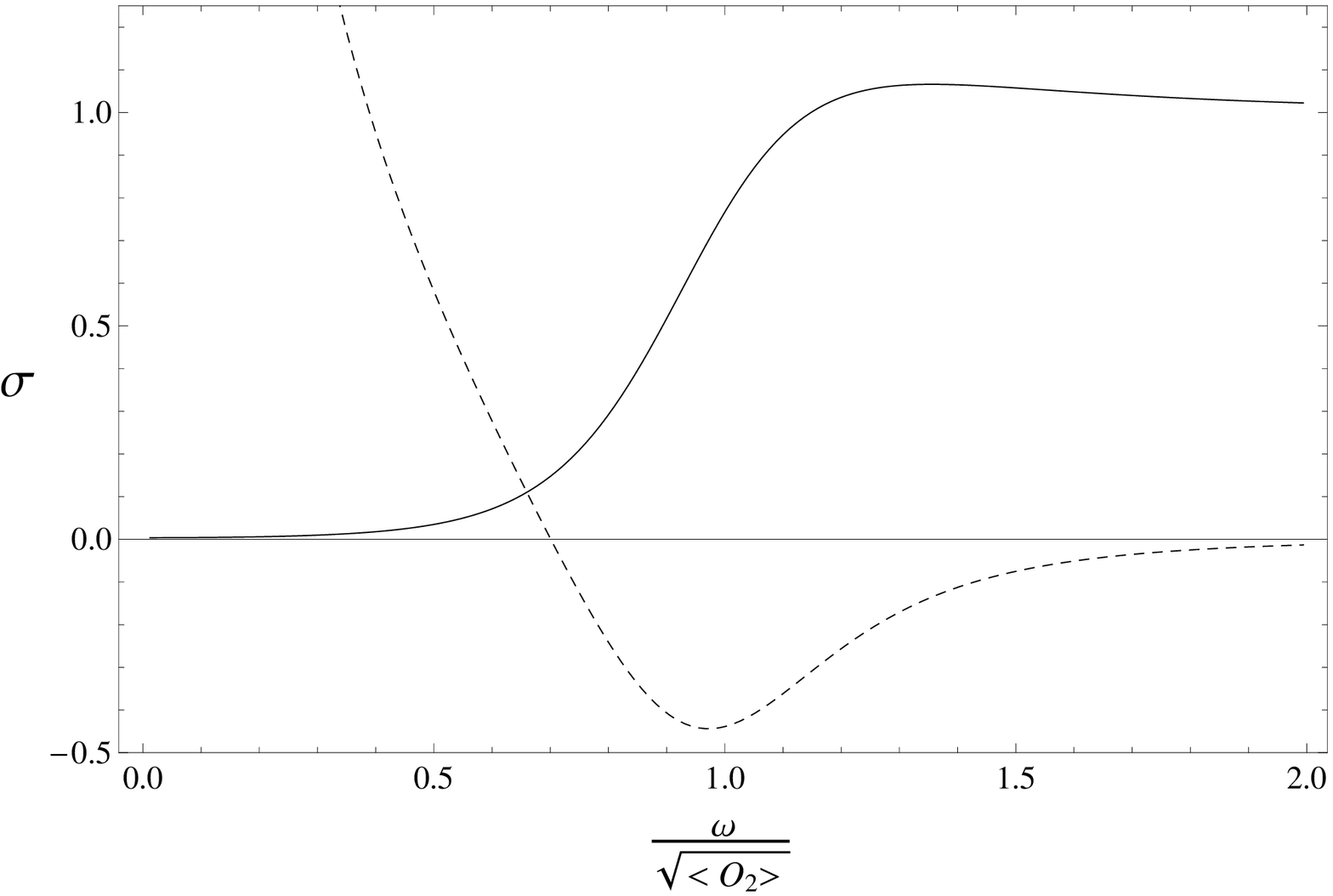}}
\subfigure{\includegraphics[width = 0.32\textwidth]{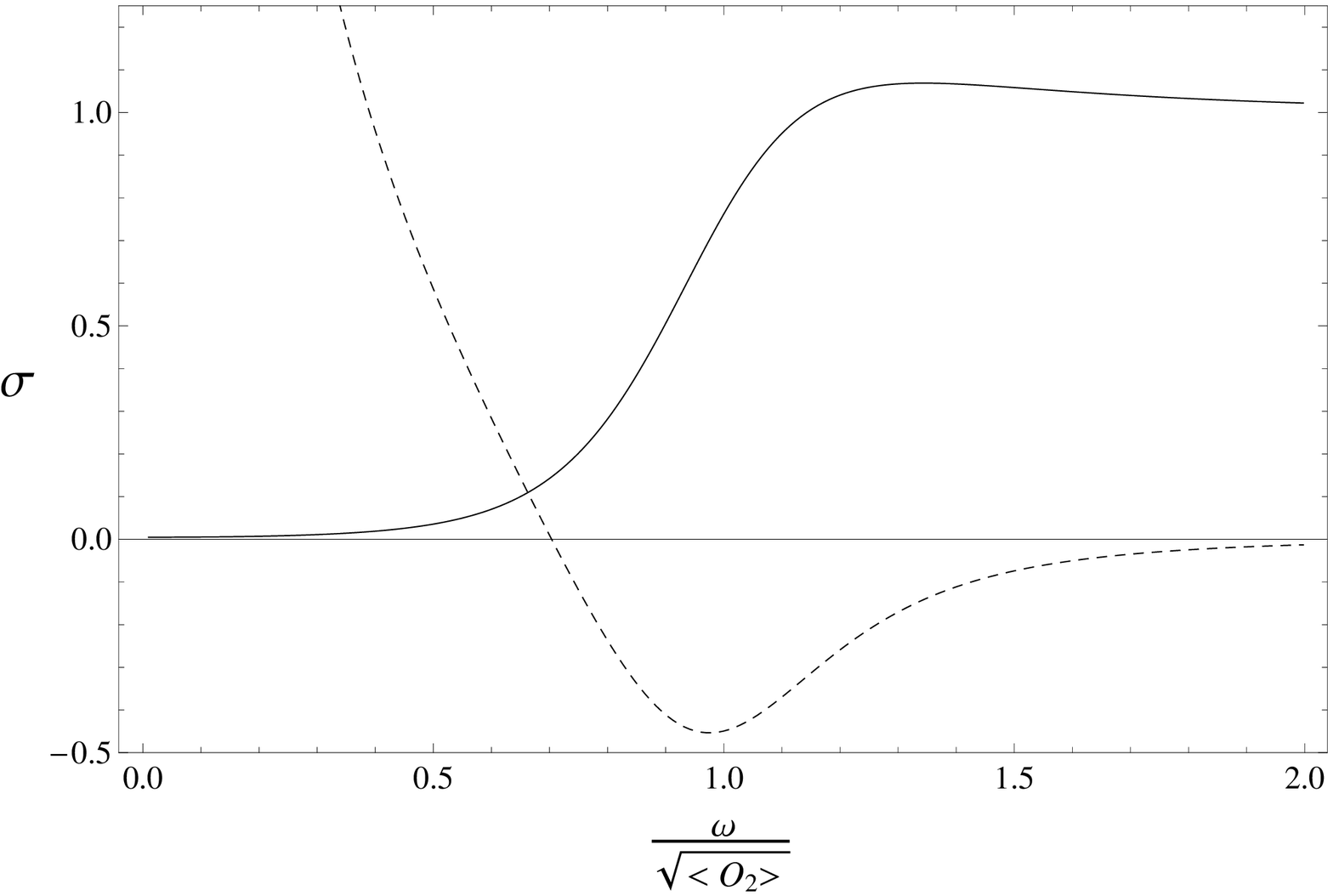}}
\subfigure{\includegraphics[width = 0.32\textwidth]{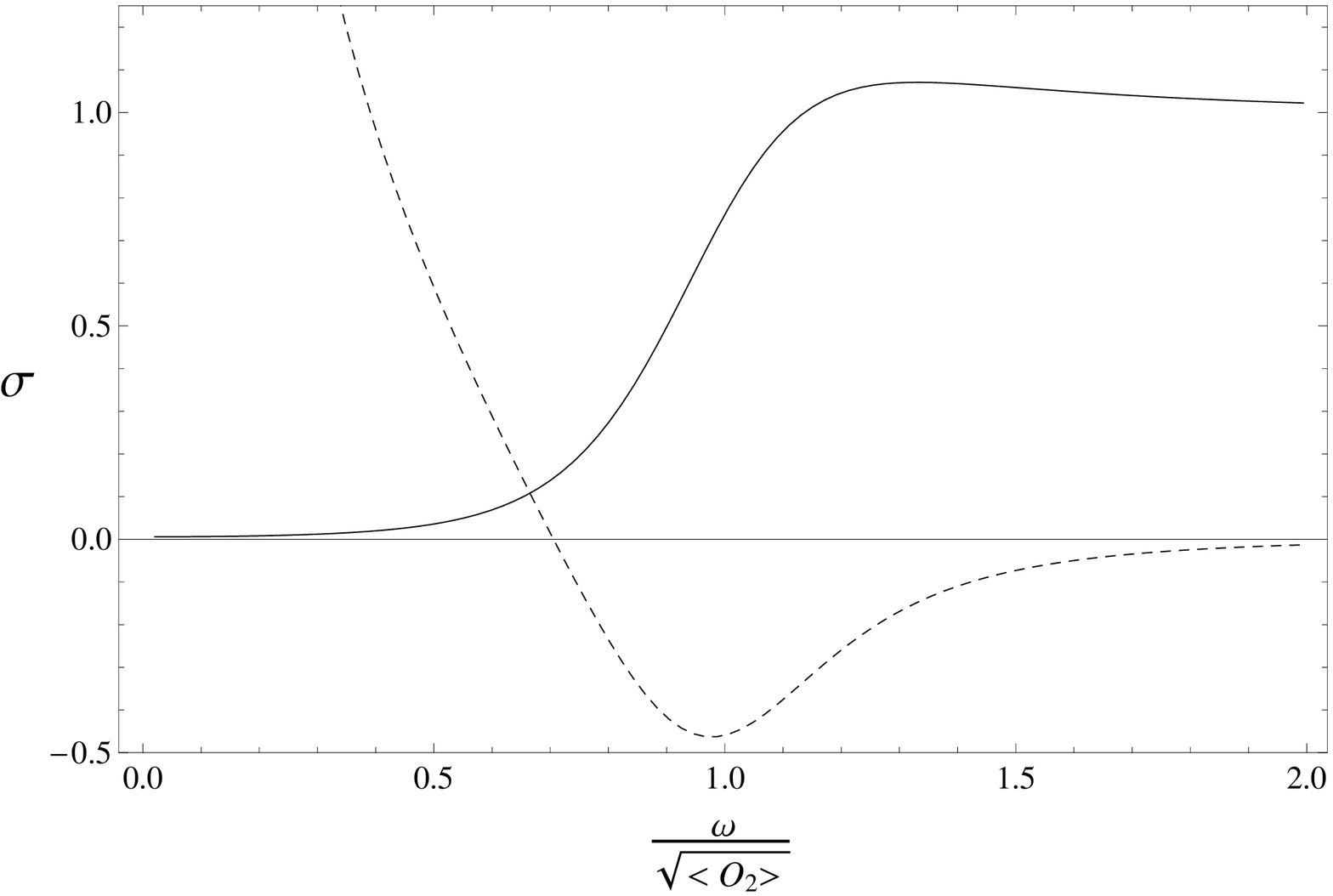}}

  \caption{
  The conductivity  at $T/T_{\rm c} = 0.5$ for the continuum model with different values of $p$ in the metric. The solid lines are the real part of the conductivity, the dashed are the imaginary part. The $p$ values for plots from left to right are $3, 3.5$ and $4$, respectively. The delta function in the real part at $\omega=0$ is not shown. Note the similarity of the three plots.}
  \label{fig:cgap}
\end{figure}

At low temperature the conductivity features a sharp gap below which the real part of the conductivity nearly vanishes.  At the gap frequency, Re$(\sigma$) display a step-function type behavior, while Im$(\sigma$) has a sharp local minimum.  Even at larger temperatures, we define the gap frequency $\omega_g$ as the
location of the local minimum of Im$(\sigma)$. In Fig.~\ref{fig:cgap}, we plot the conductivity with respect to frequency scaled in units of $\sqrt{\langle O_2\rangle}$. Note that the three plots are nearly identical. In particular, the  ratio $\omega_g/\sqrt{\langle O_2 \rangle}$  at the minimum of Im$(\sigma)$ is nearly idependent of $p$ in this range.  However,  as we will see there are important quantitative distinctions at small $\omega$.

It was noted in the original model of Ref.~\cite{HHH} that the gap-to-$T_{\rm c}$ ratio is larger in the
holographic model than in the weakly-coupled BCS theory, in rough quantitative agreement with experimental
results in high-temperature superconductors.
The normal component of the DC conductivity is defined as $n_n \equiv \lim_{\omega \to 0} {\rm Re} [\sigma(\omega)]$. For low enough temperatures, we find that \begin{equation} n_n \sim e^{- \Delta/T}\,, \label{eq:gap} \end{equation} in which $\Delta = C_{\rm p} \sqrt{\langle O_2\rangle}$, for some constant $C_p$. The coefficient $\Delta$ in the exponent is  the superconducting energy gap. In order to compare with $\omega_g$ found previously, we fit our data for a range of $T/T_{\rm c}$  around $0.5$, which gives a good exponential fit for $n_n$ in that range, with relatively large $\Delta/T>6$. 
%ranging from $0.48$ to $0.52$. 
The results are summarized in Table~I.
%NOTE: \ref{tab:cgap} gives II.

\begin{table}[ht]
\begin{center}
\begin{tabular} {|ccc ccccccc|}
\hline\hline
&$p$&&&3.0&&3.5&&4.0&\\[0.5ex]
\hline
&$\frac{\sqrt{\langle O_2\rangle}}{T_{\rm c}}$ &&& $8.28$ && $7.29$ && $6.49$&\\

&$\frac{\Delta}{ \sqrt{\langle O_2\rangle}}$ &&& $0.50$ && $0.54$ && $0.59$&\\

&$\frac{\omega_g}{ \sqrt{\langle O_2\rangle}}$ &&& $0.97$ && $0.98$ && $0.98$&\\
\hline
\end{tabular}
\caption{Observables for the continuum theory, at $T/T_{\rm c} = 0.5$.}
\end{center}
\label{tab:cgap}
\end{table}

\begin{figure}[ht]
\subfigure{\includegraphics[width = 0.495\textwidth]{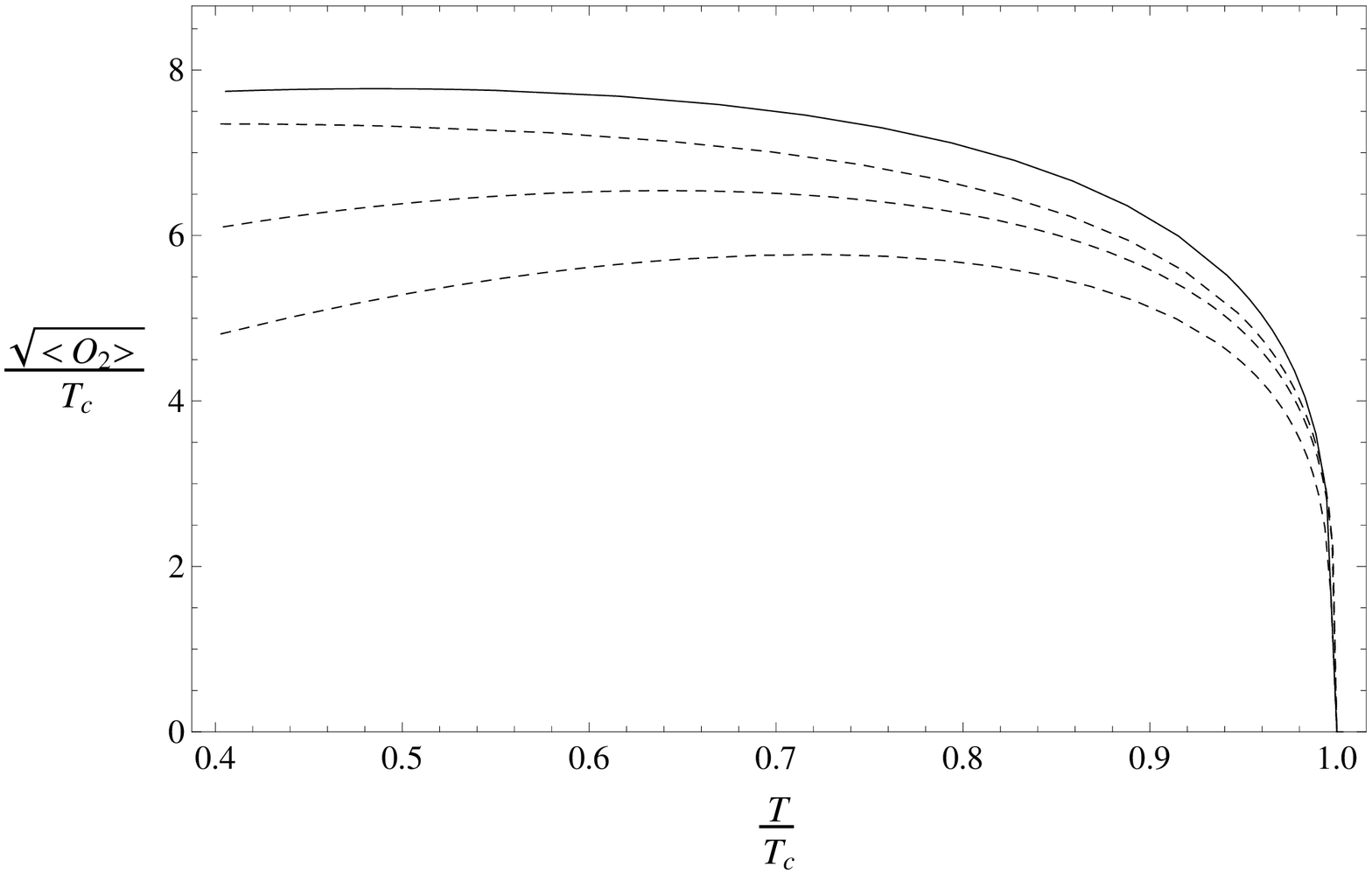}}
\subfigure{\includegraphics[width = 0.495\textwidth]{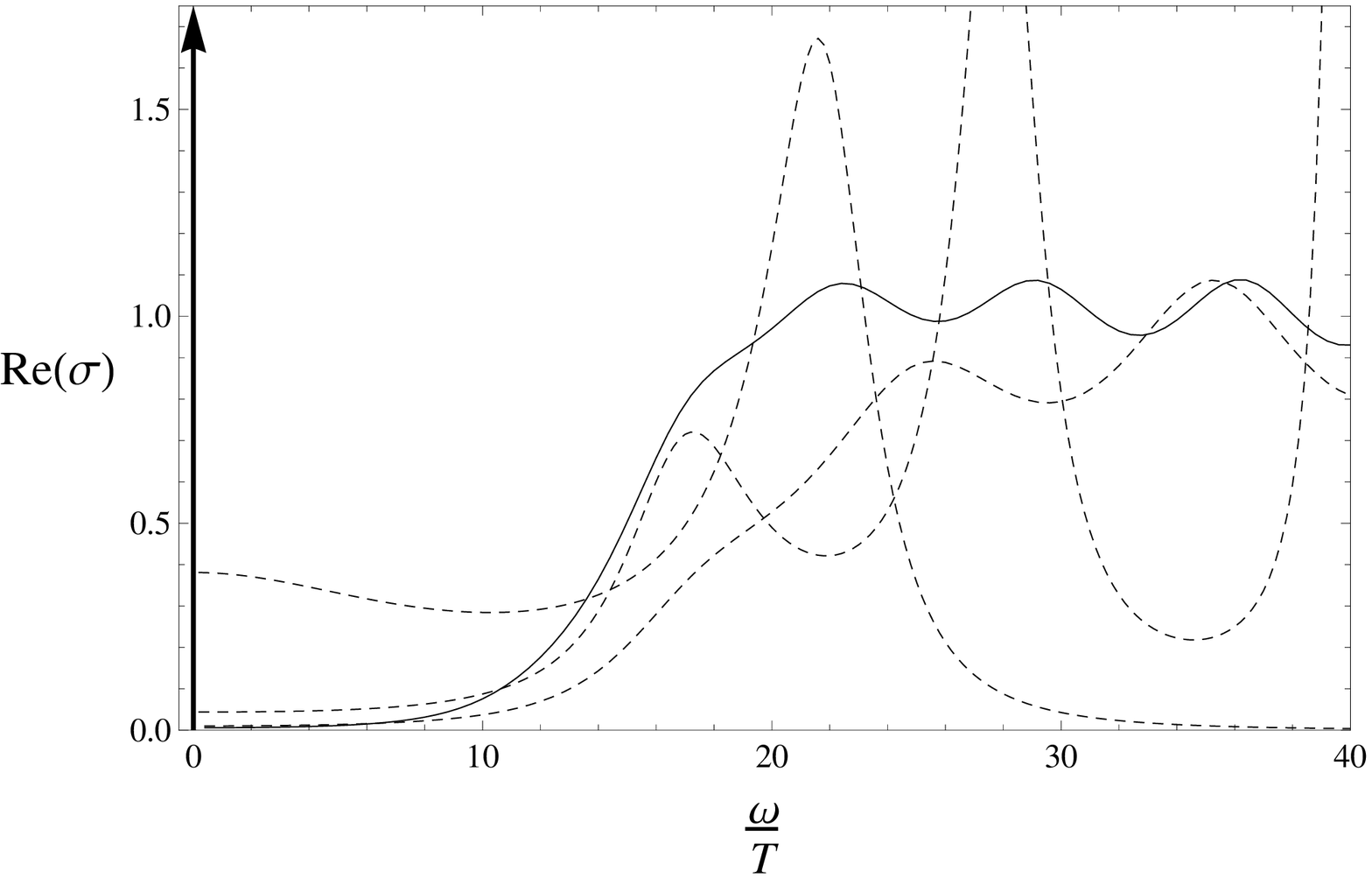}}
 \caption{
The condensate  $\langle O_{2} \rangle$ and the real part of the conductivity for the deconstructed model for $p = 3$ and $T/T_{\rm c} = 0.5$.  The solid curves correspond to $N = 1000$ lattice sites. The dashed curves, in order from top to bottom near the origin in the left-hand plot and from bottom to top in the right-hand plot correspond to $N =  100, 10$ and $5$, respectively. The arrow indicates a Dirac delta function from the  DC superconductivity.}
\label{fig:dec}
\end{figure}

\begin{figure}
\subfigure{\includegraphics[width = 0.32\textwidth]{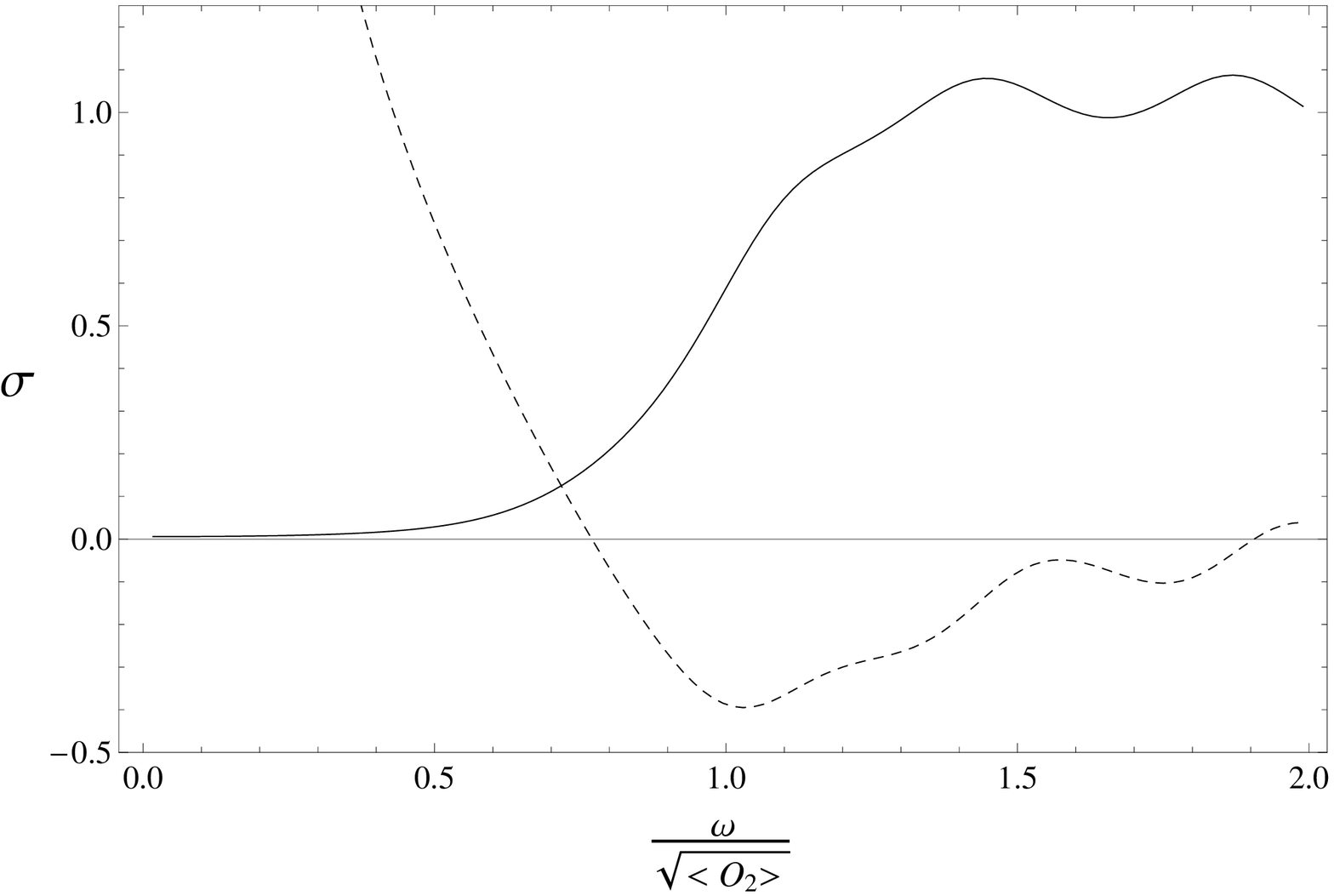}}
\subfigure{\includegraphics[width = 0.32\textwidth]{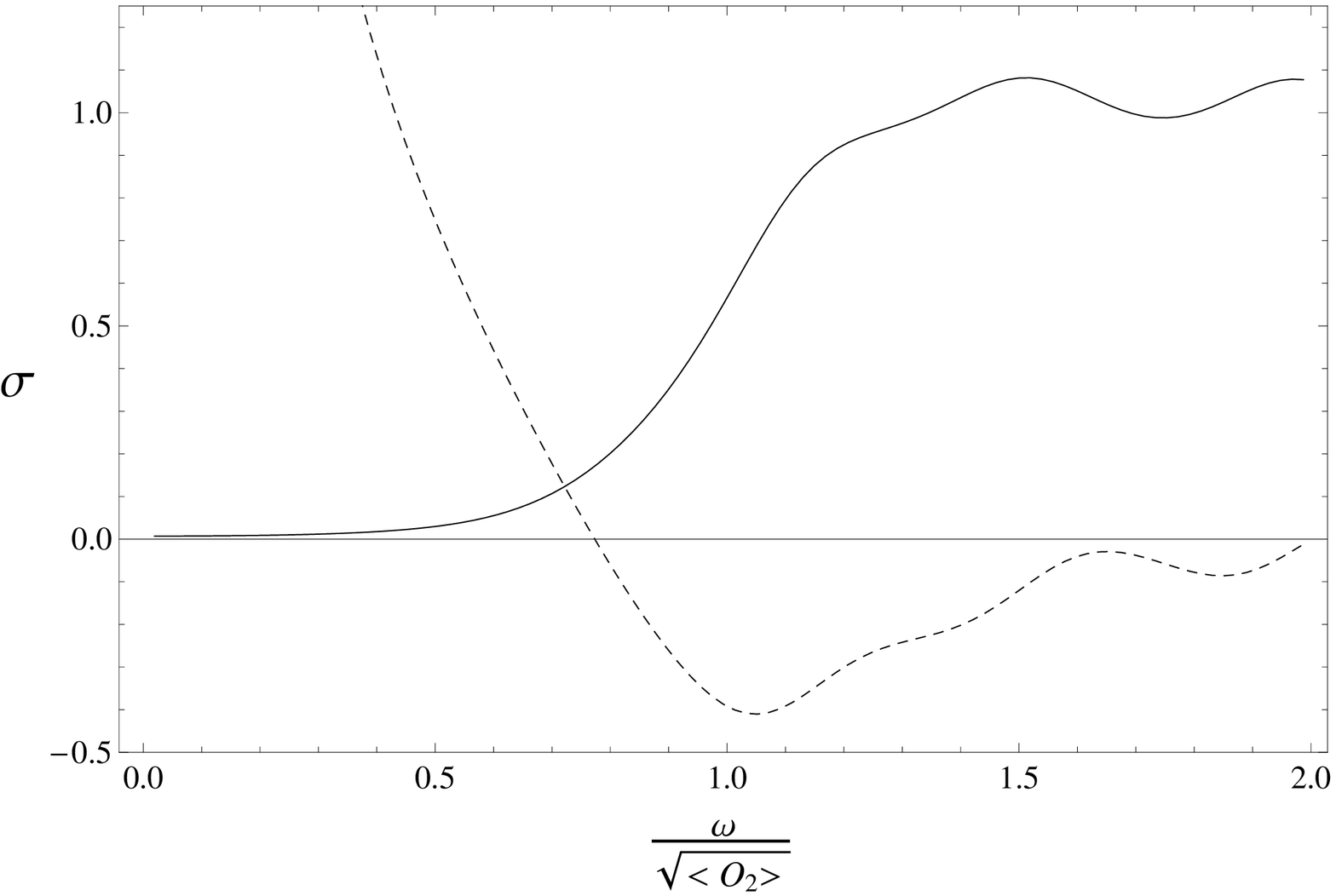}}
\subfigure{\includegraphics[width = 0.32\textwidth]{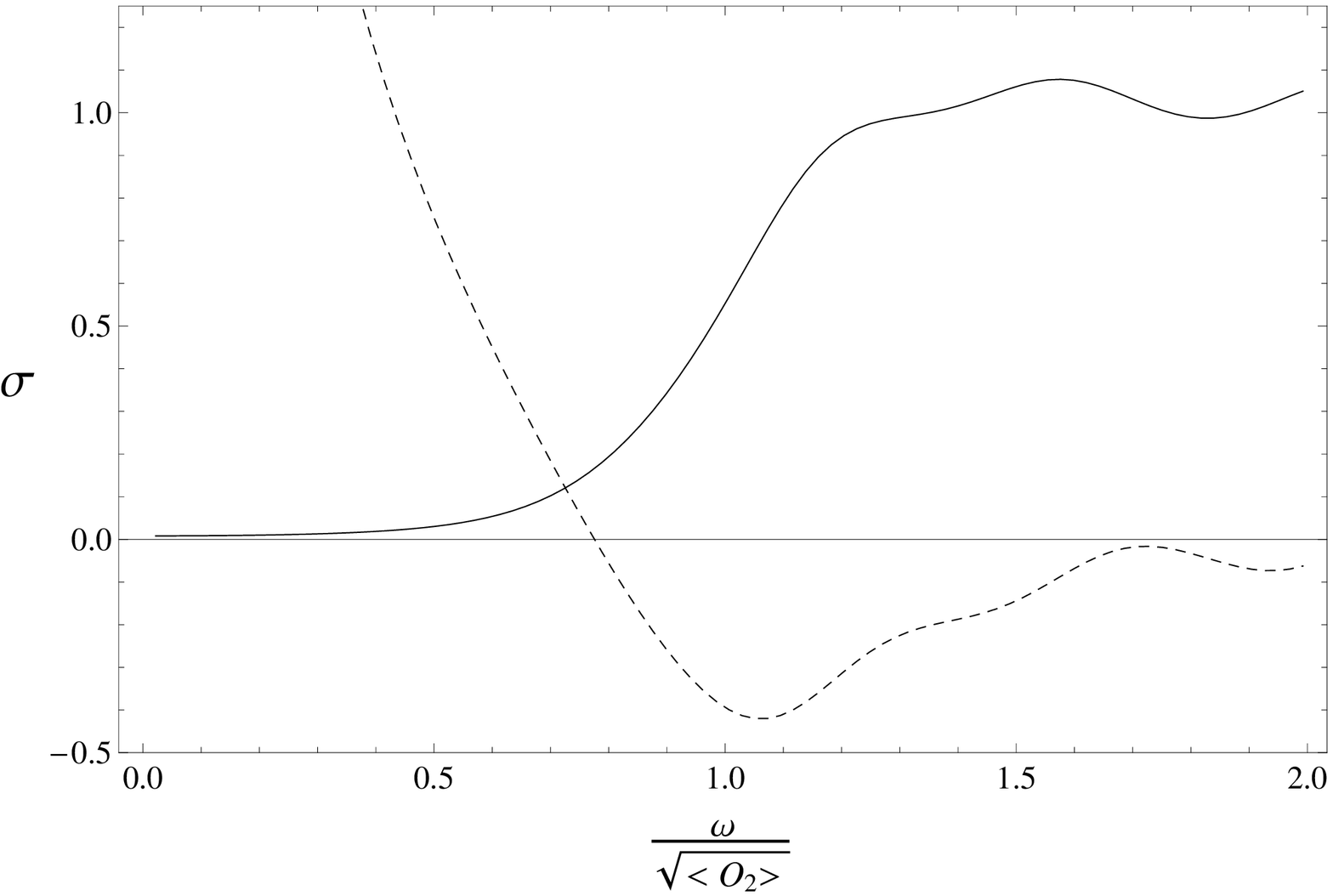}}
\subfigure{\includegraphics[width = 0.32\textwidth]{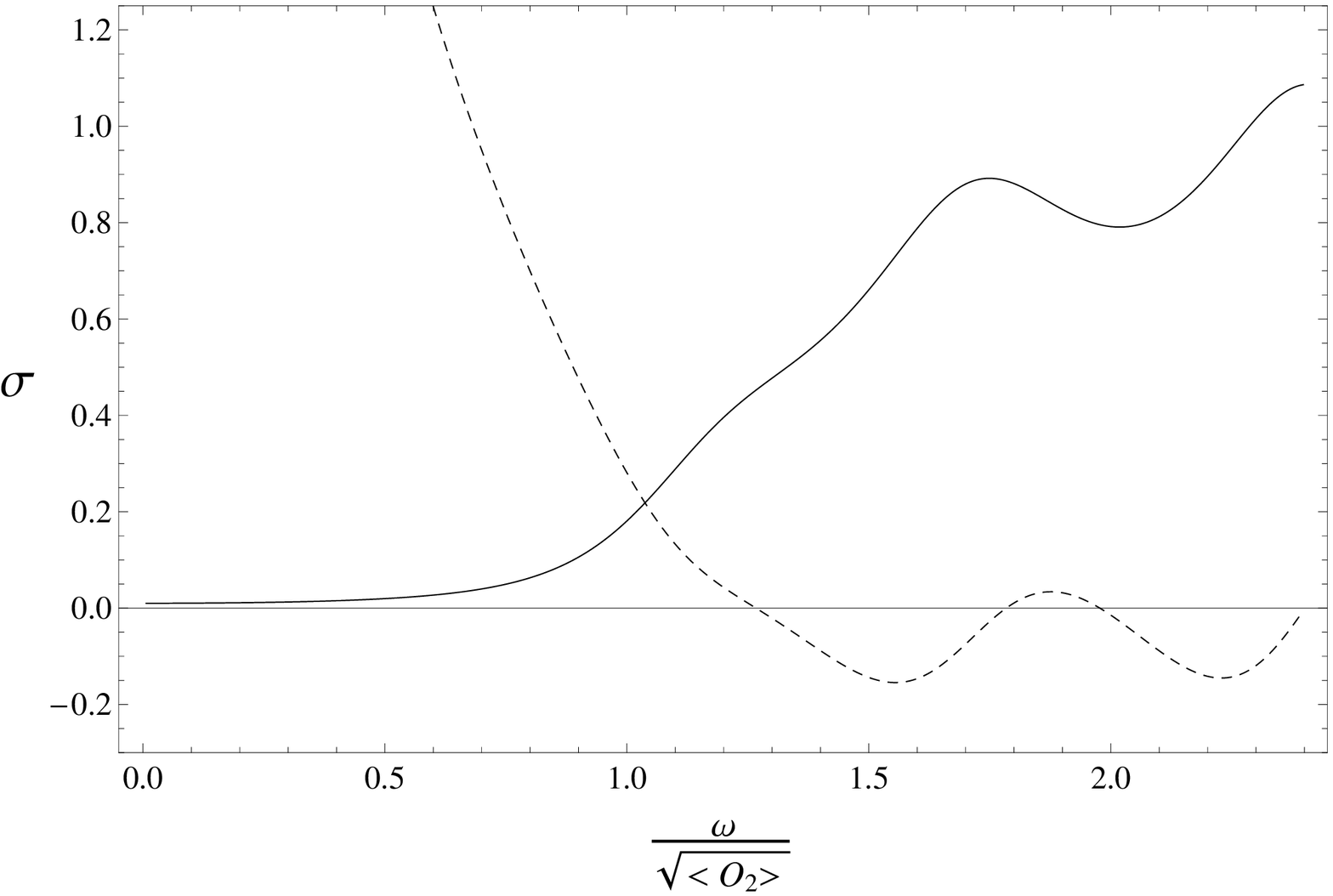}}
\subfigure{\includegraphics[width = 0.32\textwidth]{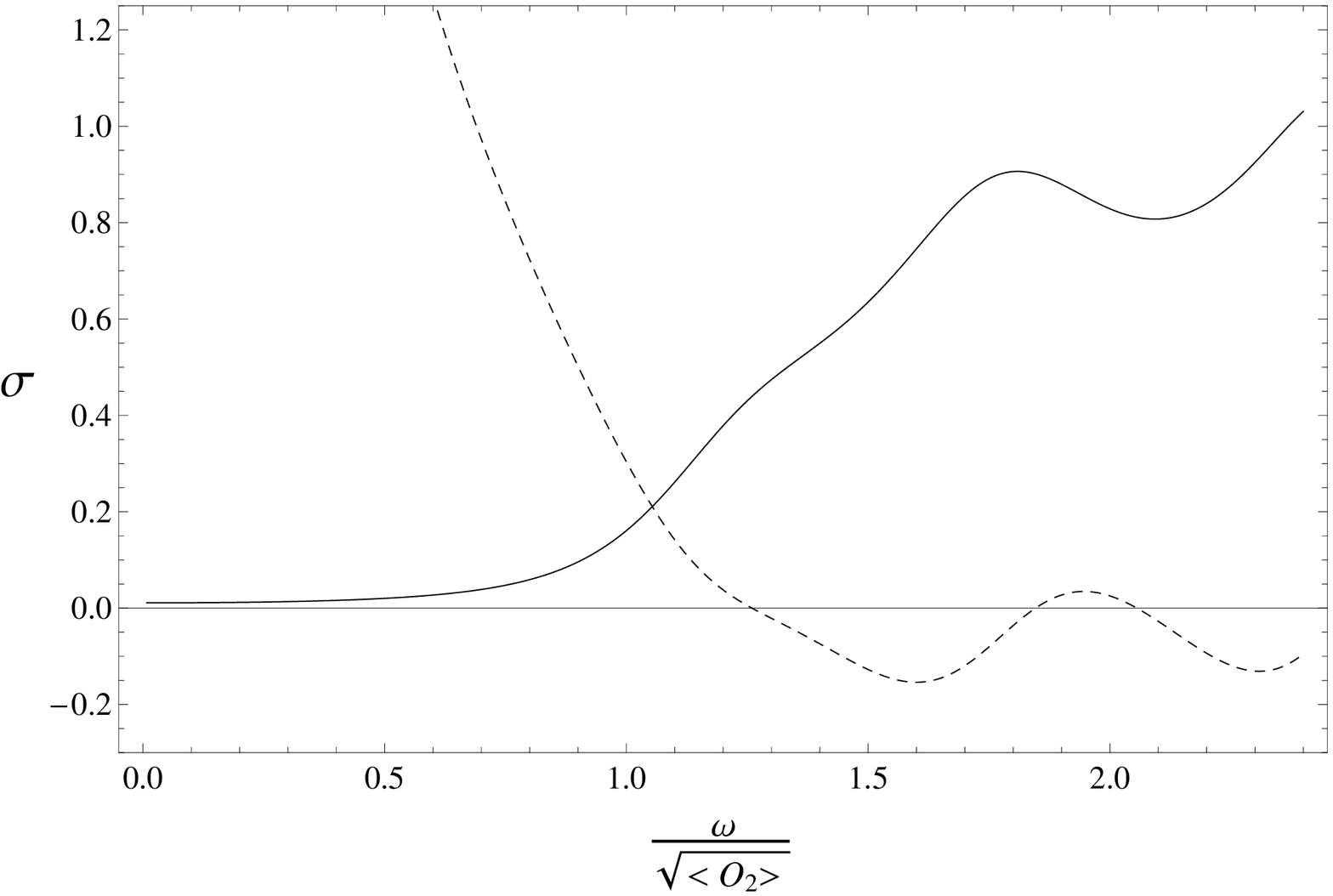}}
\subfigure{\includegraphics[width = 0.32\textwidth]{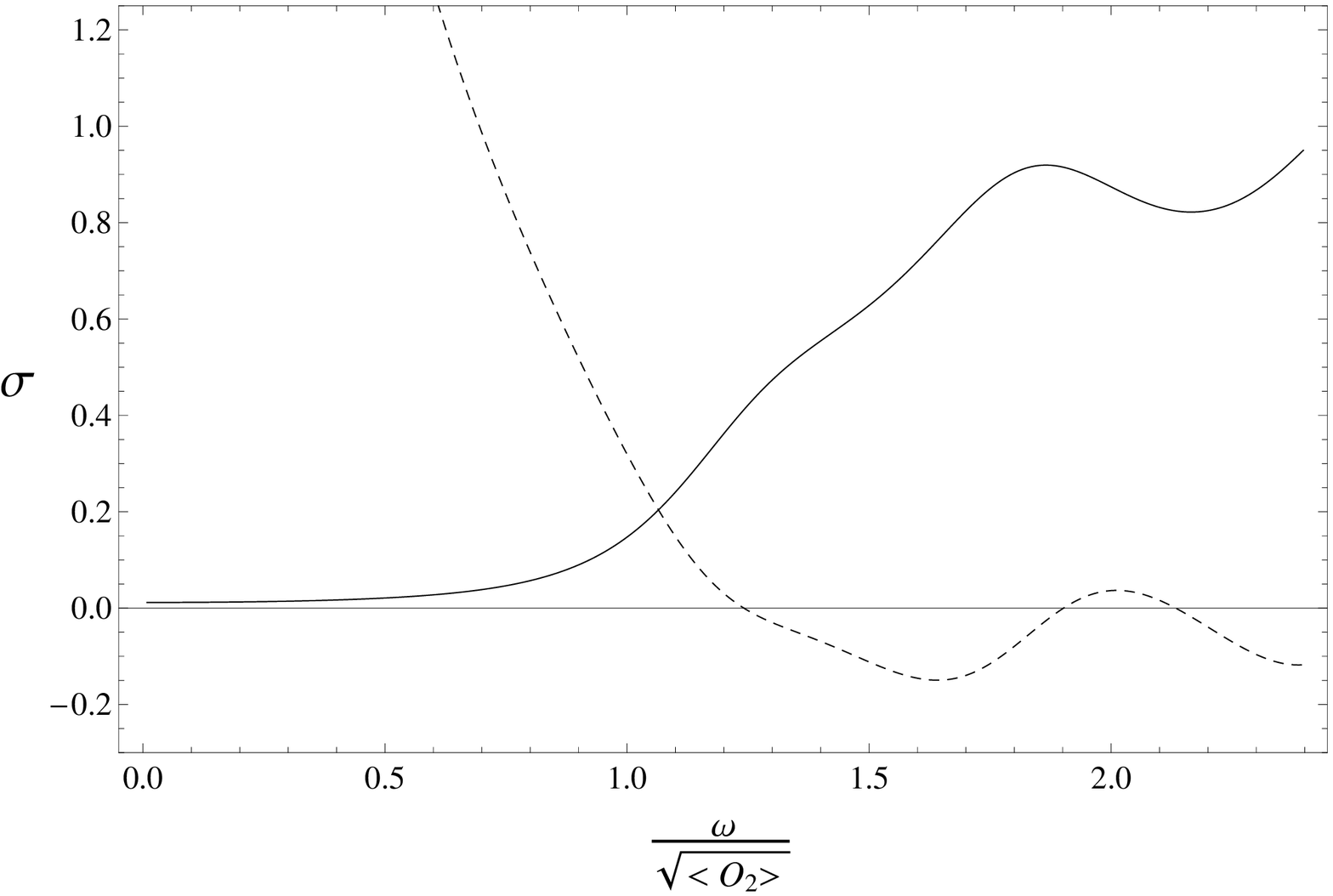}}
\subfigure{\includegraphics[width = 0.32\textwidth]{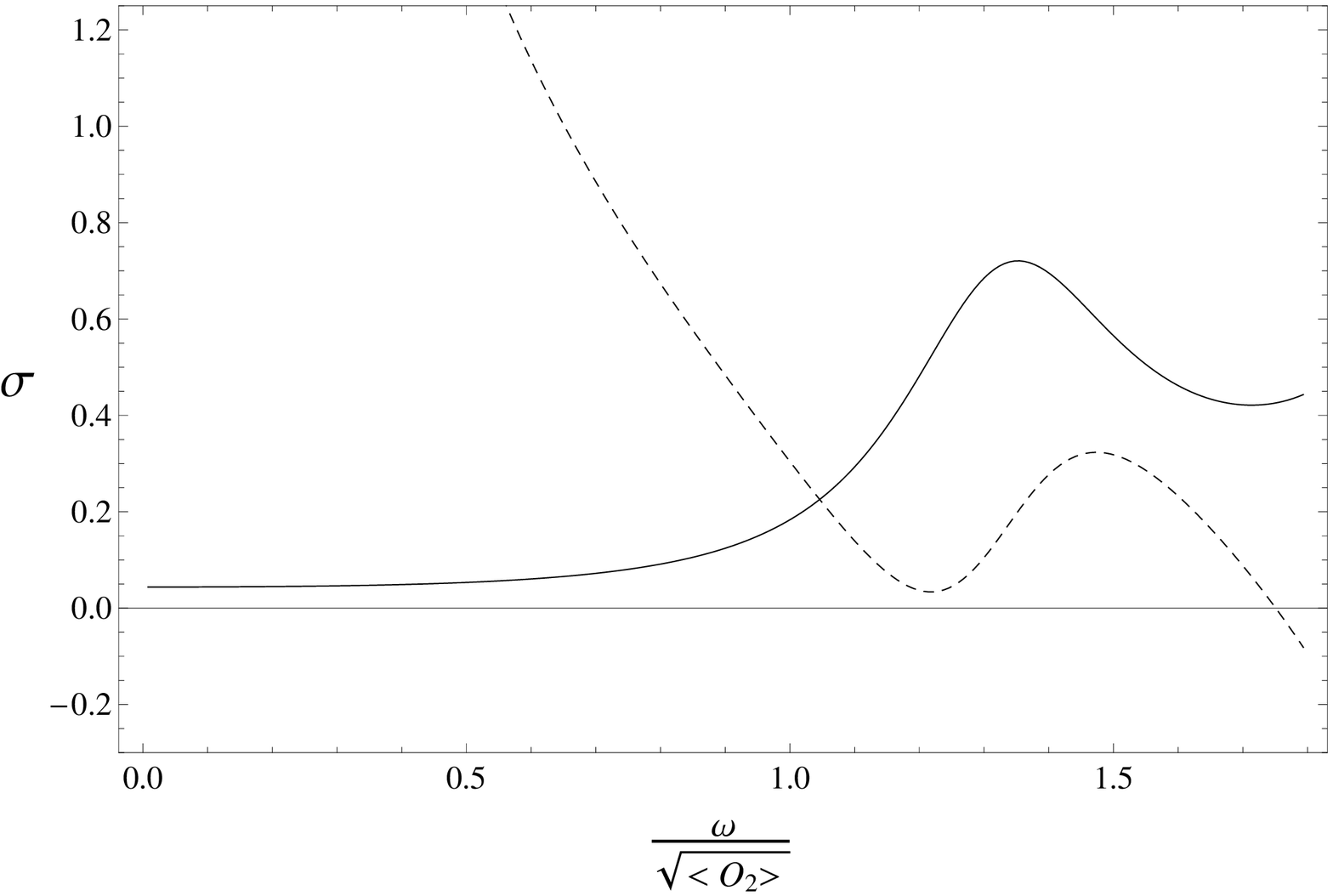}}
\subfigure{\includegraphics[width = 0.32\textwidth]{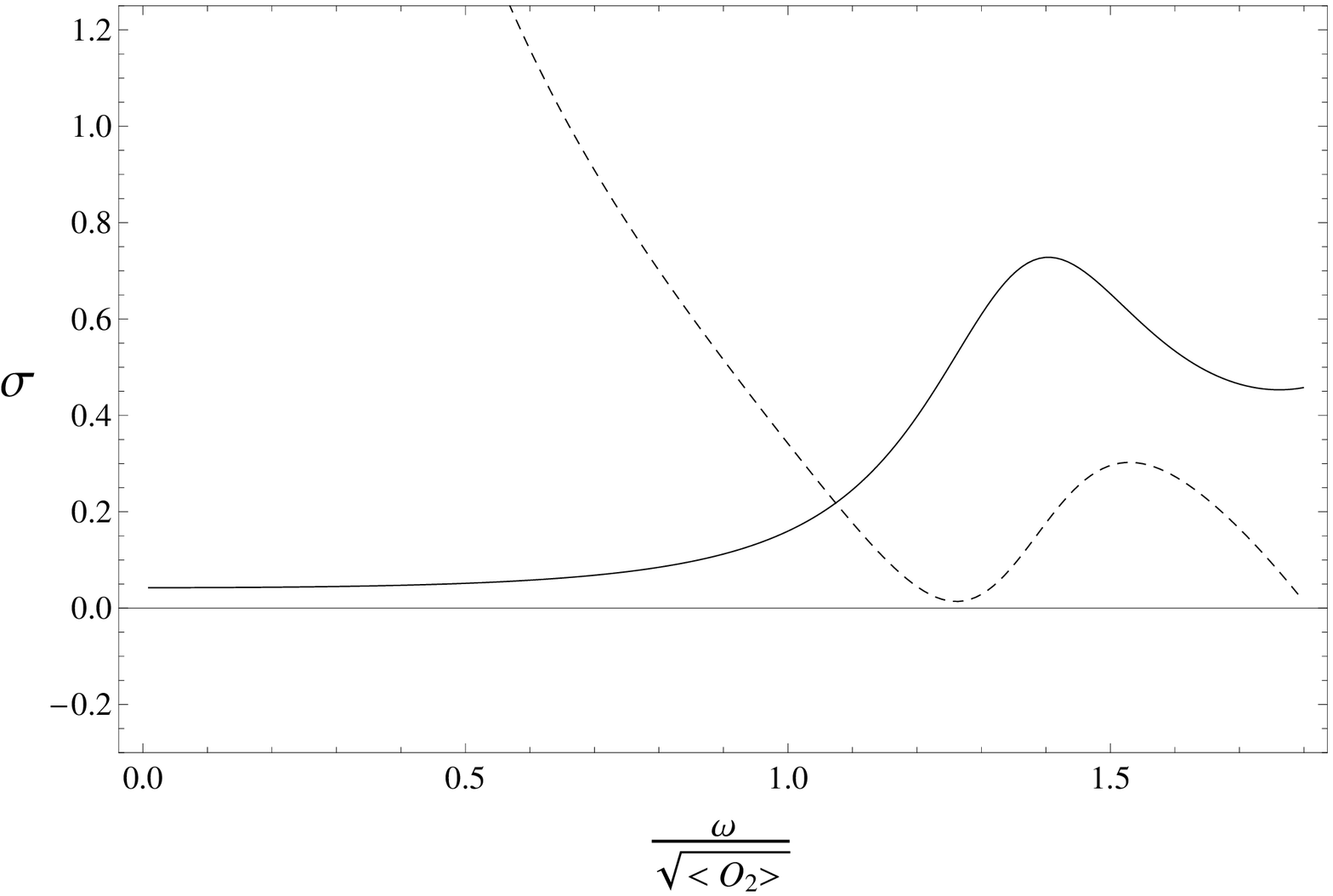}}
\subfigure{\includegraphics[width = 0.32\textwidth]{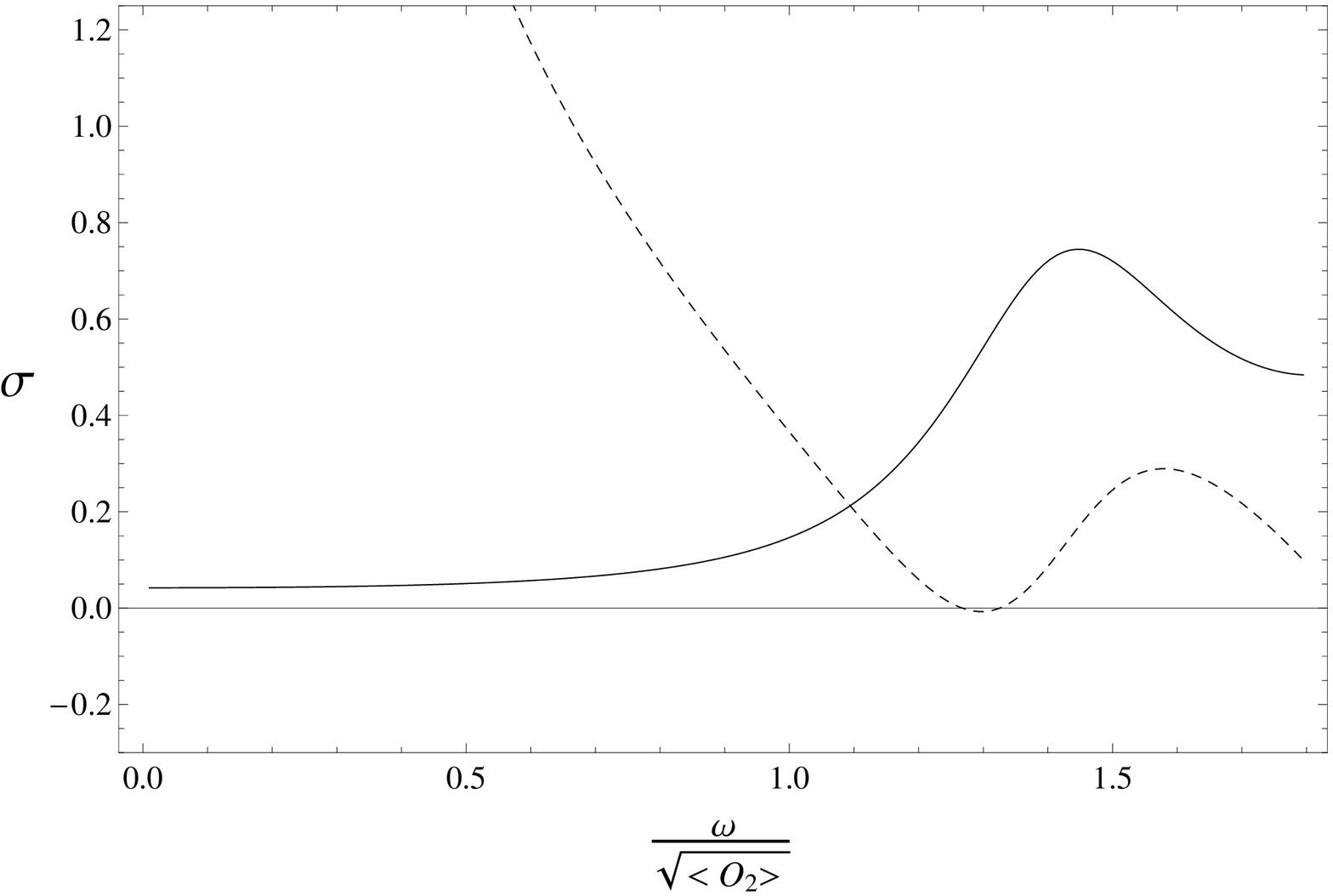}}
\subfigure{\includegraphics[width = 0.32\textwidth]{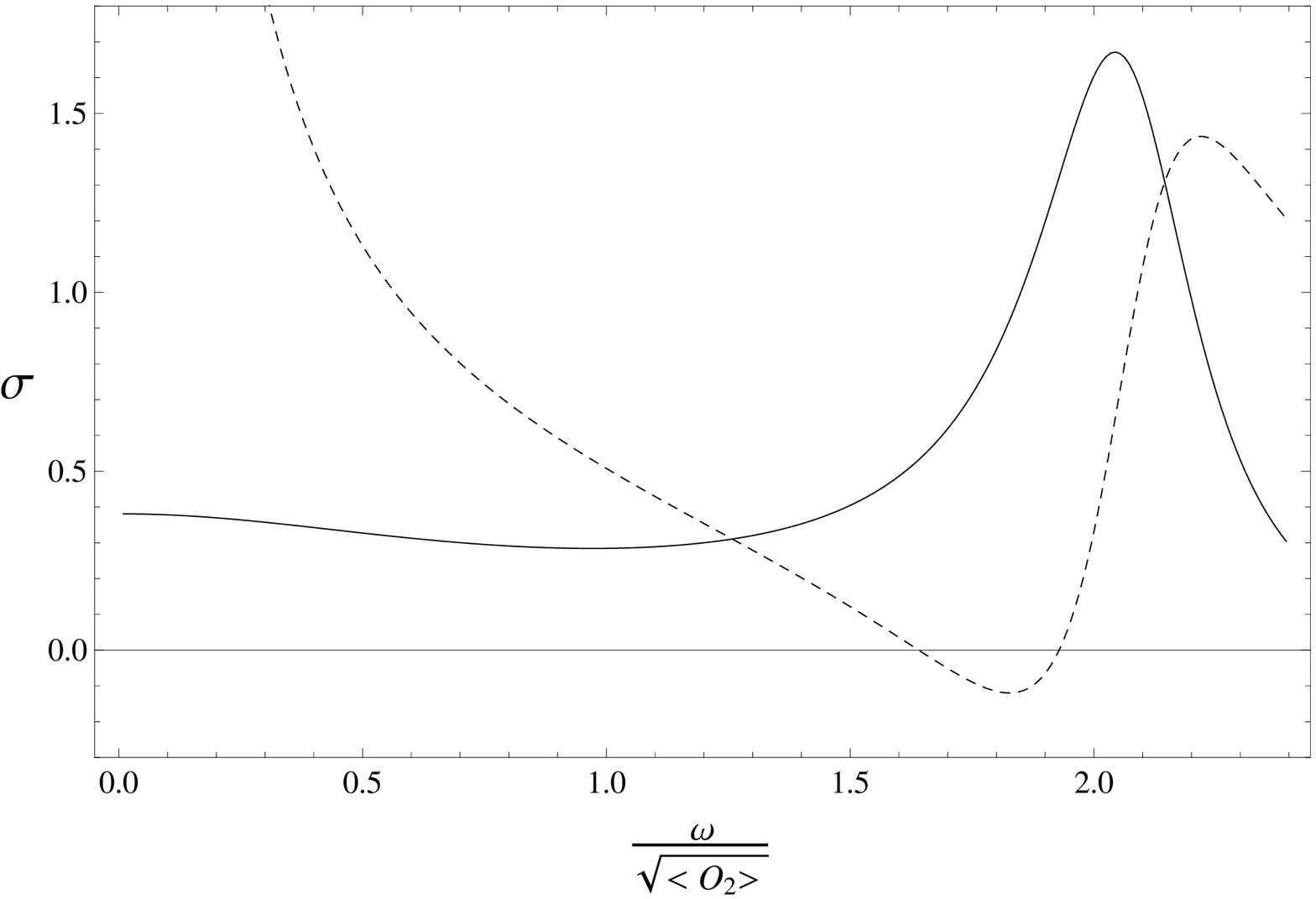}}
\subfigure{\includegraphics[width = 0.32\textwidth]{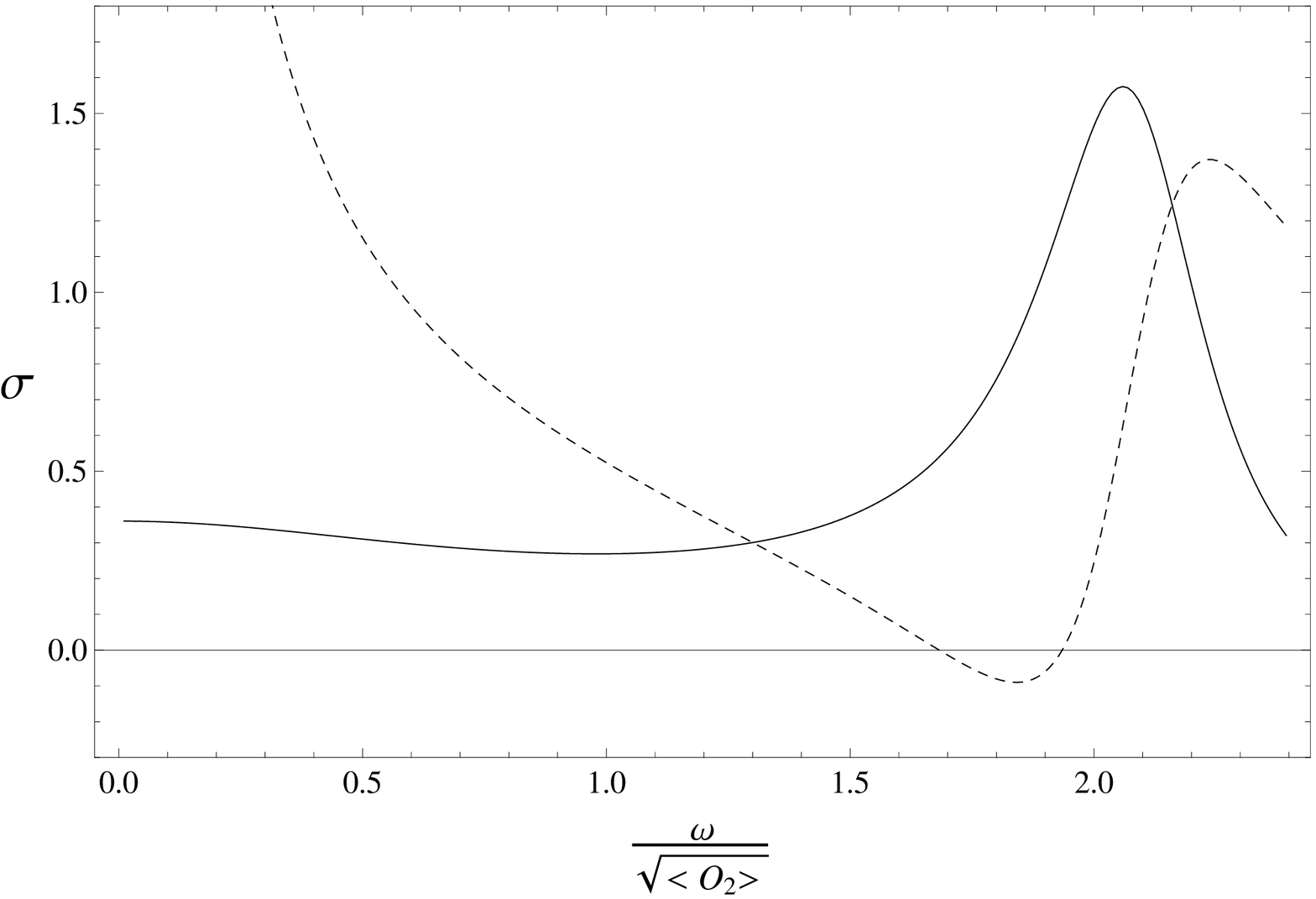}}
\subfigure{\includegraphics[width = 0.32\textwidth]{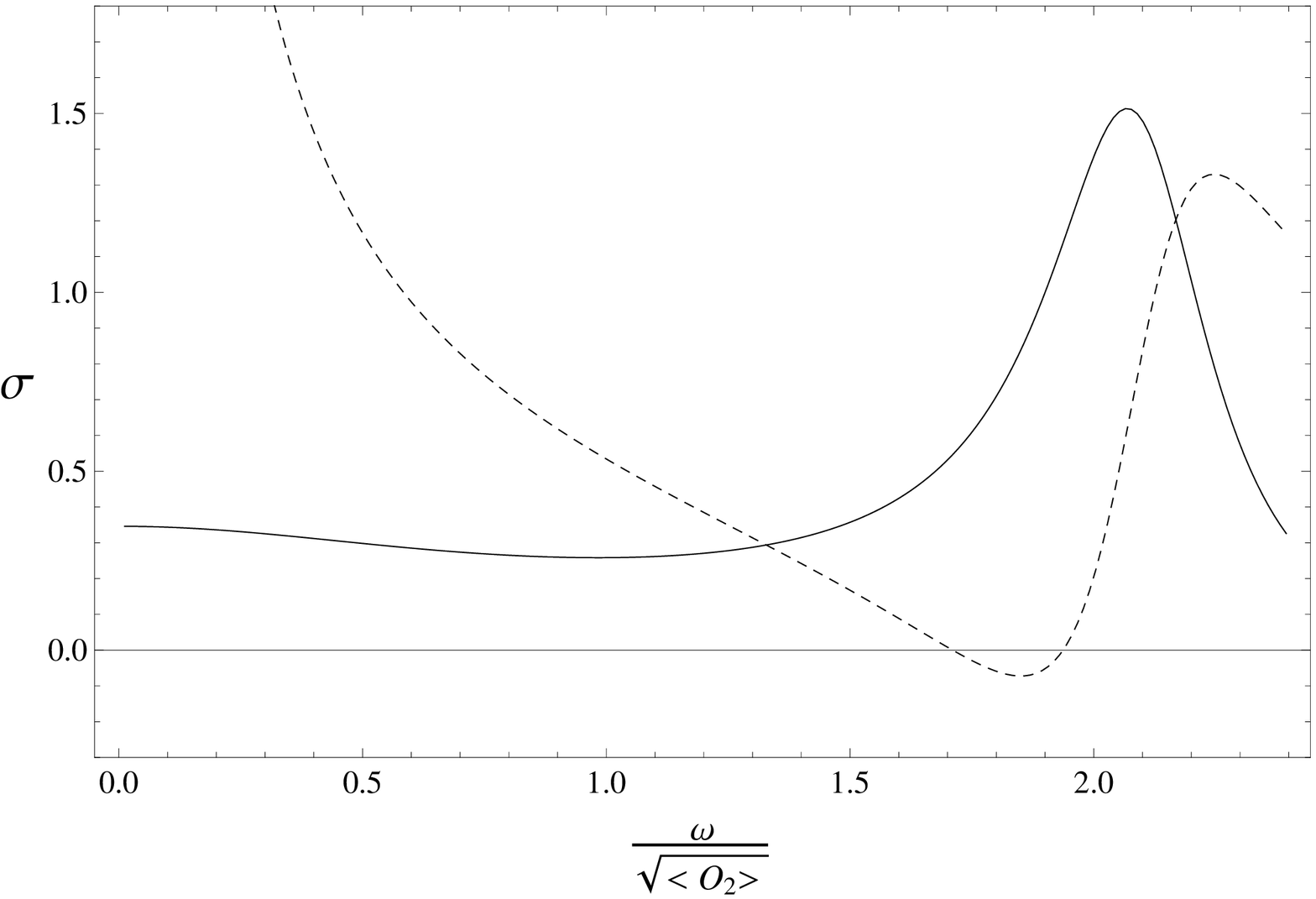}}
  \caption{
  The conductivity at $T/T_{\rm c} = 0.5$ for the deconstructed model. The solid lines are the real part of conductivity, the dashed are imaginary. The $p$ values for plots from left to right are $3, 3.5$ and $4$, respectively. The rows of plots from top to bottom correspond to $N = 1000, 100, 10$ and $5$, respectively.}
  \label{fig:gap}
\end{figure}

We next examine the deconstructed model with $p = 3, 3.5$ and $4$ for $N \in \left\{5, 10, 100, 1000\right\}$. We generally set the UV cutoff at $z=\epsilon = 1$, except for the case $N=1000$, for which we set $\epsilon= 0.1$  to better match the continuum model. The lattice spacing at the horizon is fixed at $a_H = 10^{-5}$, decoupled from the lattice spacing in the bulk which varies as the horizon moves with temperature. We again use a scaling relation to set $\rho = 1$ so that $T_{\rm c}$ is fixed. As discussed in \cite{ACE}, we have the following discretized version of the boundary conditions:
\begin{equation}
\phi_{1}' = -\rho = -1\, ,\quad \psi^{(1)} = 0\, ,\quad
  \phi_{N} = 0\, ,\quad {\rm and} \,\,
  \psi_{N-1}' = \frac{2}{3 z_{N}} \psi_{N},
\end{equation}
where the primes are discretized derivatives as in Eq.~\ref{eq:deriv}.
Electromagnetism is defined as the U(1) interaction at the UV boundary site, {\em i.e.} the lattice site closest to $z=0$.  We find solutions for which the $x$-component of the bulk gauge fields oscillate while the other components do not fluctuate, \begin{equation}
A_{xi}(t)=e^{-i\omega t} A_{xi}, \label{eq:Axi} \end{equation}
where on the right-hand side of Eq.~(\ref{eq:Axi}), $A_{xi}$ is time-independent.
The conductivity $\sigma = J^x_{1}/E_{x1}$ is found to be given by a discretized version of the 
holographic calculation for $\sigma$ in the continuum model:
\begin{equation}
\sigma = - \frac{ i f_1(A_{x2} - A_{x1}) / a}{\omega A_{x1}}\,.
\end{equation}
To obtain reasonable phenomenology we find that an ingoing-wave type boundary condition is necessary even in the deconstructed models.  Due to the behavior of the metric near the horizon, we find it beneficial to impose a discretized version of the ingoing-wave boundary condition a bit away from the horizon in order to better mimic the continuum solutions.  In particular, we impose the frequency-dependent boundary condition of Ref.~\cite{ACE}:
\begin{equation}
A_{x\, N - n} = 1 \quad {\rm and} \quad A_{x\, N - n - 1} = 1 -
  \frac{i \omega a}{f_{N - n - 1}} \, .
\end{equation}  The shift into the bulk, given by  $n$, is chosen to be $n = 20, 10, 2$ and $2$ for $N = 1000, 100, 10$ and $5$, respectively. In Fig.~\ref{fig:dec} we plot the  condensate and the real part of the conductivity for $p = 3$. It was suggested in Ref.~\cite{ACE} that the large resonances in the conductivity may correspond to exciton-polariton resonances due to the broken U(1) gauge groups in the model. The $p = 3.5$ and $4$ cases are qualitatively similar, and some examples are given in Fig.~\ref{fig:gap}. The critcal temperatures at which the condensate starts to form are listed in Table~\ref{tab:tc}.

\begin{table}[ht]
\begin{center}
\begin{tabular} {|cc crccccccccc|}
\hline\hline
&$N$ && &&1000&&100&&10&&5&\\[0.5ex]
\hline
&&& $p = 3$ && $0.118$ && $0.104$ && $0.094$ && $0.079$&\\
&$T_{\rm c}$ && $3.5$ && $0.134$ && $0.118$ && $0.107$ && $0.090$&\\
&&& $4$ && $0.151$ && $0.132$ && $0.121$ && $0.101$&\\
\hline
\end{tabular}
\caption{Critical temperatures in units of $\rho^{1/2}$.}
\end{center}
\label{tab:tc}
\end{table}

To further analyze observables in the deconstructed models we mimic the analysis of the continuum model. It can be seen directly from the locations of the minimum of Im$(\sigma)$ in Fig.~\ref{fig:gap} that $\omega_g/\sqrt{\langle O_{2}\rangle}\neq 1$, but its value is not sensitive to $p$ in the range examined, even in the 5-site model. The relation (\ref{eq:gap}) continues to be well satisfied and defines the gap $\Delta$ as in the continuum model.  The pole in the imaginary part of the conductivity is manifest in Fig.~\ref{fig:gap}, and is related to the delta-function in the real part via a Kramers-Kronig relation.  The results for observables are listed in Table~III.
%NOTE: \ref{tab:dgap} gives II. 

\begin{table}[htbp]
\begin{center}
\begin{tabular} {|ccc rccccccccc|}
\hline\hline
&$N$ && &&1000&&100&&10&&5&\\[0.5ex]
\hline
&&& $p = 3$ && $7.77$ && $7.32$ && $6.38$ && $5.29$&\\
&$\frac{\sqrt{\langle O_2\rangle}}{T_{\rm c}}$ && $3.5$ && $6.85$ && $6.48$ && $5.61$ && $4.62$&\\
&&& $4$ && $6.09$ && $5.78$ && $4.99$ && $4.09$&\\[1ex]
\hline
&&& $3$ && $0.52$ && $0.52$ && $0.48$ && $0.28$&\\
&$\frac{\Delta}{\sqrt{\langle O_2\rangle}}$ && $3.5$ && $0.58$ && $0.58$ && $0.56$ && $0.33$&\\
&&& $4$ && $0.64$ && $0.64$ && $0.63$ && $0.38$&\\[1ex]
\hline
&&& $3$ && $1.03$ && $1.55$ && $1.22$ && $1.83$&\\
&$\frac{\omega_g}{\sqrt{\langle O_2\rangle}}$ && $3.5$ && $1.05$ && $1.60$ && $1.26$ && $1.84$&\\
&&& $4$ && $1.06$ && $1.64$ && $1.30$ && $1.85$&\\
\hline
\end{tabular}
\caption{Observables for the deconstructed model. The $\sqrt{\langle O_2\rangle}$ and $\omega_g$ in the table are taken at $T/T_{\rm c} = 0.5$.}
\end{center}
\label{tab:dgap}
\end{table}

\section{Conclusions}
We have analyzed the dependence of the charged condensate and the complex conductivity on the form of the black-hole metric in holographic superconductors and in deconstructed versions of those models.  We found that certain model predictions are relatively insensitive to the details of the spacetime.  For example, the approximate relation between the gap frequency  and the superconducting condensate,
\begin{equation}
\omega_g/\sqrt{\langle O_2\rangle}= 1, \end{equation} 
persists while the metric is varied in the continuum model.  In the deconstructed model this ratio differs from 1, but remains insensitive to the deconstructed metric.  On the other hand, we have seen relatively strong dependence of other observables on the details of the metric, such as the ratio of the superconducting gap $\Delta$ to $T_{\rm c}$.  Furthermore, in deconstructed models we found that this ratio can be significantly smaller than in the continuum model. The model sensitivity supports the conclusion that the quantitative success of the simplest holographic models of superconductors is accidental.  However, qualitative features of these and related models continue to suggest the possibility of explaining some of the unusual properties of unconventional superconductors.  For example, in addition to a large gap,  related models describe a strange metallic phase
\cite{strange-metal} in which the resistivity does not vary like $T^2$ as in the Fermi liquid description of metals.  In an effort to make contact with physical systems, it remains important to continue to investigate  which aspects of the holographic models and their deconstructed cousins are responsible for the nonconventional behavior of the superconductors described by these models.

\section{Acknowledgments}
We are grateful to Patrick King for collaboration at early stages of this work.  We also thank Dylan Albrecht and Chris Carone for useful communication.  This work was supported by the NSF under Grant PHY-1068008.  JE also thanks Edward Coco for his generous support.

\end{document}